\documentclass{aa}

\usepackage{graphicx,natbib,txfonts,color}

\newcommand{\msun}{$M_{\odot}$} 

\def\al#1{\mbox{$^{#1}$Al}}

\def\car#1{\mbox{$^{#1}$C}}

\def\mg#1{\mbox{$^{#1}$Mg}}

\def\nit#1{\mbox{$^{#1}$N}}

\def\ox#1{\mbox{$^{#1}$O}}

\begin{document}

   \title{On the asymptotic giant branch star origin \\ of peculiar spinel grain OC2} 
\subtitle{}

   \author{M. Lugaro\inst{1}
          \and
          A. I. Karakas\inst{2}
          \and
          L. R. Nittler\inst{3}
          \and\\
          C. M. O'D. Alexander\inst{3}
          \and
          P. Hoppe\inst{4}
          \and
          C. Iliadis\inst{5}
          \and
          J. C. Lattanzio\inst{6}
          }

   \offprints{M. Lugaro, M.Lugaro@phys.uu.nl}

   \institute{Sterrenkundig Instituut, University of Utrecht, Postbus 80000, 
         3508 TA Utrecht, The Netherlands
         \and
         Origins Institute, Department of Physics \& Astronomy, 
         McMaster University, Hamilton ON, Canada
         \and
         Carnegie Institution of Washington, Department of 
         Terrestrial Magnetism Washington DC 20015, USA
         \and
         Max-Planck-Institute for Chemistry, P.O. Box 3060, 55020 Mainz, Germany
         \and
         Department of Physics and Astronomy, 176 Phillips Hall, University of North Carolina, 
         Chapel Hill, NC 27599-3255, USA
         \and
         Centre for Stellar and Planetary Astrophysics, School of Mathematical Sciences, Monash
         University, Victoria 3800, Australia
         }

   \date{Received June --, 2006; accepted -- --, 2006}

   \authorrunning{M. Lugaro et al.}
   \titlerunning{Peculiar spinel grain OC2}

\abstract
{Microscopic presolar grains extracted from primitive meteorites have extremely anomalous isotopic 
compositions revealing the stellar origin of these grains. Multiple elements in single presolar 
grains can be analysed with sensitive mass spectrometers, providing precise sets of isotopic 
compositions to be matched by theoretical models of stellar evolution and nucleosynthesis.}
{The composition of presolar spinel grain OC2 is different from that of all other presolar spinel 
grains. In particular, large excesses of the heavy Mg isotopes are present and thus an origin from an 
intermediate-mass (IM) asymptotic giant branch (AGB) star was previously proposed for this grain. 
We discuss the O, Mg, Al, Cr and Fe isotopic compositions of presolar spinel grain OC2 and compare them 
to theoretical predictions.}
{We use detailed models of the evolution and nucleosynthesis of AGB stars
of different masses and metallicities to compare to the composition of grain OC2. We analyse the 
uncertainties related to nuclear reaction rates and 
also discuss stellar model uncertainties.}
{We show that the isotopic composition of O, Mg and Al in OC2 could be the signature of an AGB 
star of IM and metallicity close to solar experiencing hot bottom burning, or of an AGB star of 
low mass (LM) and low metallicity ($\simeq$0.004) suffering very efficient cool bottom 
processing. Large measurement uncertainty in the Fe isotopic composition prevents us from discriminating 
which model better represents the parent star of OC2. However, the Cr isotopic composition of the 
grain favors an origin in an IM-AGB star of metallicity close to solar.} 
{Our IM-AGB models produce a self-consistent solution to match 
the composition of OC2 within the uncertainties related to reaction rates. Within this solution we 
predict that the $^{16}$O($p,\gamma)^{17}$F and the $^{17}$O($p,\alpha)^{14}$N reaction rates should be 
close to their lower and upper limits, respectively. 
By finding more grains like OC2 and by precisely measuring their Fe and Cr isotopic compositions,   
it may be possible in the future to derive constraints on massive AGB models from the study of 
presolar grains.}


\maketitle

\section{Introduction} \label{sec:intro}

Presolar grains were born in circumstellar regions around ancient stars, ejected into the 
interstellar medium, preserved during the formation of the Solar System, and trapped inside primitive 
meteorites from which they are now extracted and analysed by various microanalytical techniques. 
Their isotopic compositions are extremely anomalous compared to those found in the bulk of materials 
formed in the Solar System (e.g., Earth, meteorites, etc.). They represent a detailed record of 
the compositions of their parent stars, and, as such, provide major constraints for models of stellar 
structure and nucleosynthesis and of the chemical evolution of the Galaxy  
\citep{zinner:98,clayton:04,lugaro:05}. 
Until recently, Mg-rich presolar grains have been thought to be rare, as the 
dominant identified carbide (SiC, graphite) and oxide (corundum: Al$_2$O$_3$) phases have low Mg 
contents. 
The situation changed with the advent of a new type of ion microprobe with improved 
spatial resolution and sensitivity, the Cameca NanoSIMS 50 \citep{hillion:94}.  Using the 
NanoSIMS, 
\citet{zinner:03} found that presolar spinel is, in fact, more abundant in meteorites than is 
presolar corundum, but has a finer grain size, with most grains having diameters less 
than 1 $\mu$m.


\begin{figure}
\resizebox*{\hsize}{!}{\includegraphics{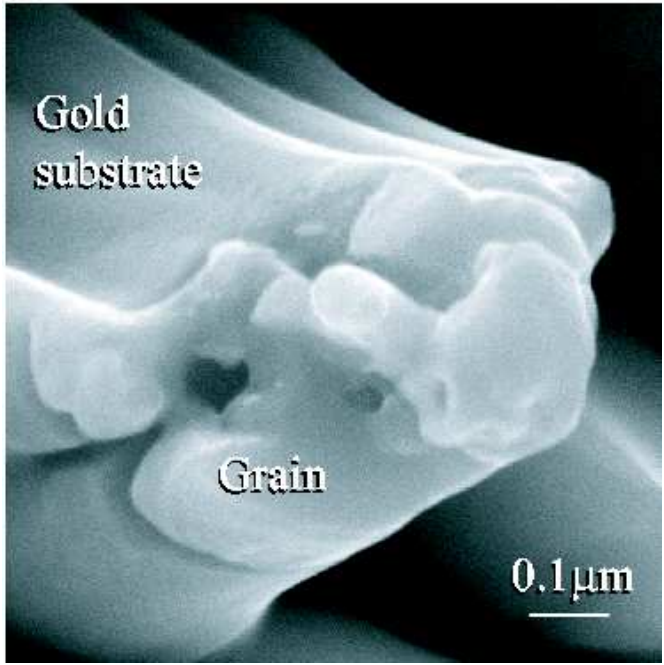}}
\caption{Scanning electron microscope image of presolar spinel grain OC2. This 800nm 
grain is sitting on a gold pedestal, following the ion probe isotopic analysis, because the gold 
substrate sputters faster than the grain does.}
\label{fig:oc2pic}
\end{figure}

\begin{figure}
\resizebox*{\hsize}{!}{\includegraphics{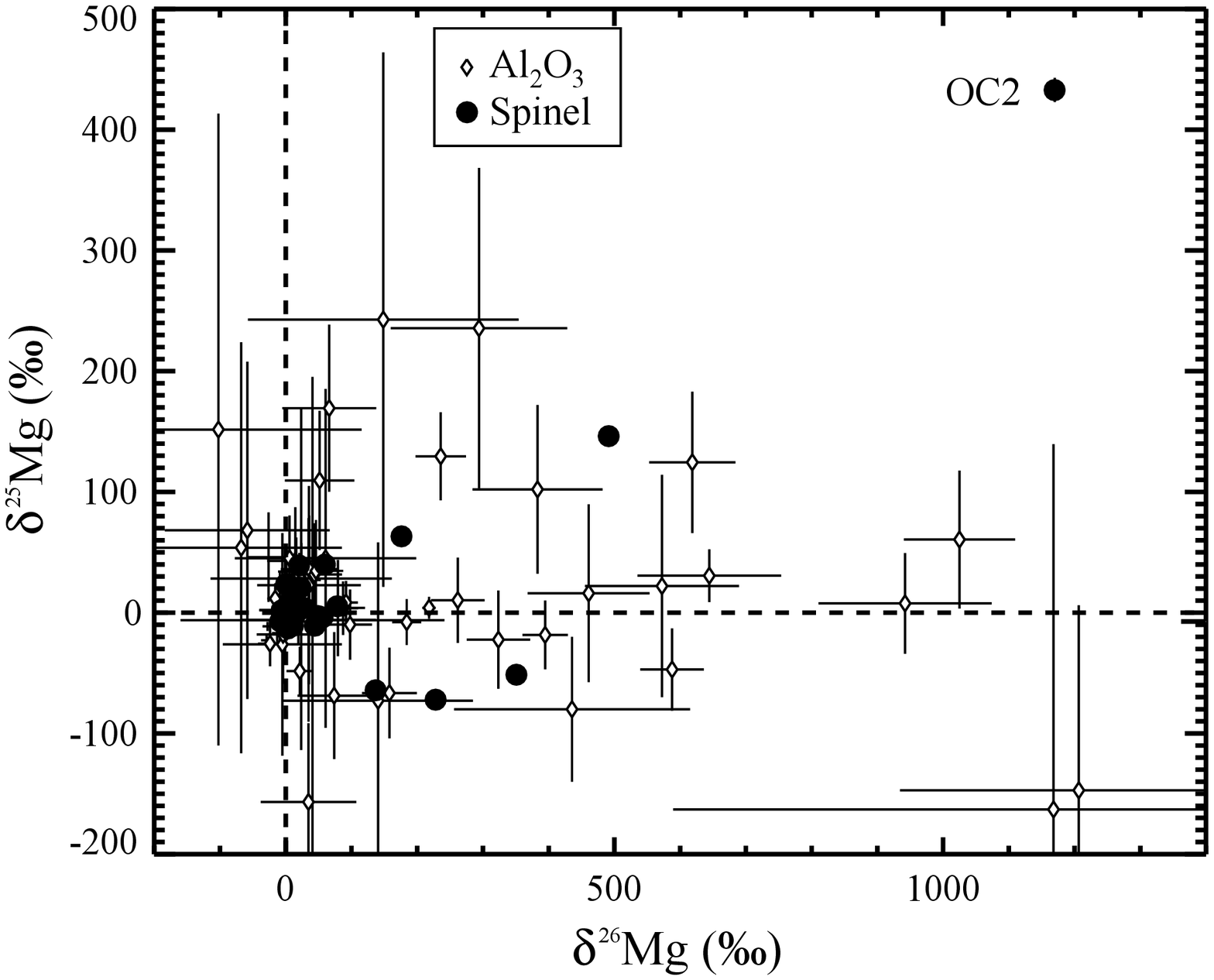}}
\caption{Mg isotopes in presolar spinel and corundum grains 
\citep{nittler:97,choi:99,zinner:05}. Dashed lines indicate solar isotopic ratios; errors are 
1-$\sigma$. Spinel grain 
OC2 has larger excesses of \mg{25} and \mg{26} than seen in other presolar spinel grains. Measured 
$\delta(^{26}$Mg/$^{24}$Mg) values for corundum extend up to 1.7$\times$10$^{6}$, due to high initial 
contents of 
\al{26} and high Al/Mg ratios, but no corundum grain has a \mg{25} enrichment similar to that of OC2.} 
\label{fig:grains_mg}
\end{figure}

A presentation and discussion of the properties and the compositions of the more than 
300 presolar spinel grains reported so far can be found in \citet{zinner:05} and \citet{nguyen:03}.
Most presolar oxide as well as silicate grains show enhancements in $^{17}$O and 
depletions in $^{18}$O similarly to what is observed in red giant and low-mass (LM) AGB stars, pointing to 
such an origin for the majority of these grains \citep{nittler:97}. These compositions are explained by 
the effect 
of the first dredge-up and extra-mixing processes.
We focus in this paper on the composition of a single extraordinary presolar spinel grain, 
named OC2. OC2, of size $\approx$0.8 $\mu$m (Fig.~\ref{fig:oc2pic}), was identified in a mixed acid 
residue of the Semarkona, Krymka and Bishunpur unequilibrated ordinary chondrites 
\citep{nittler:99,zinner:05}. 
{The isotopic composition of this grain (Table~\ref{tab:OC2})} is peculiar among presolar oxide grains. 
Its most remarkable 
feature are large excesses of the heavy Mg isotopes: $^{25}$Mg and $^{26}$Mg are enriched with respect 
to solar composition by 43\% and 117\%, respectively (Fig.~\ref{fig:grains_mg}). The Mg isotopic 
composition is represented in Table~\ref{tab:OC2} by the $\delta$ notation:  $$ \delta(^{i}{\rm 
Mg}/^{24}{\rm Mg})= \bigg(\frac{(^{i}{\rm Mg}/^{24}{\rm Mg})_{measured}} {(^{i}{\rm Mg}/^{24}{\rm 
Mg})_{solar}} 
-1\bigg) \times 1000, $$ which we will use throughout this paper,
where the abundant $^{24}$Mg, which represents about 79\% of all magnesium 
in the Solar System, is used as the reference isotope. 

\begin{figure}
\resizebox*{\hsize}{!}{\includegraphics{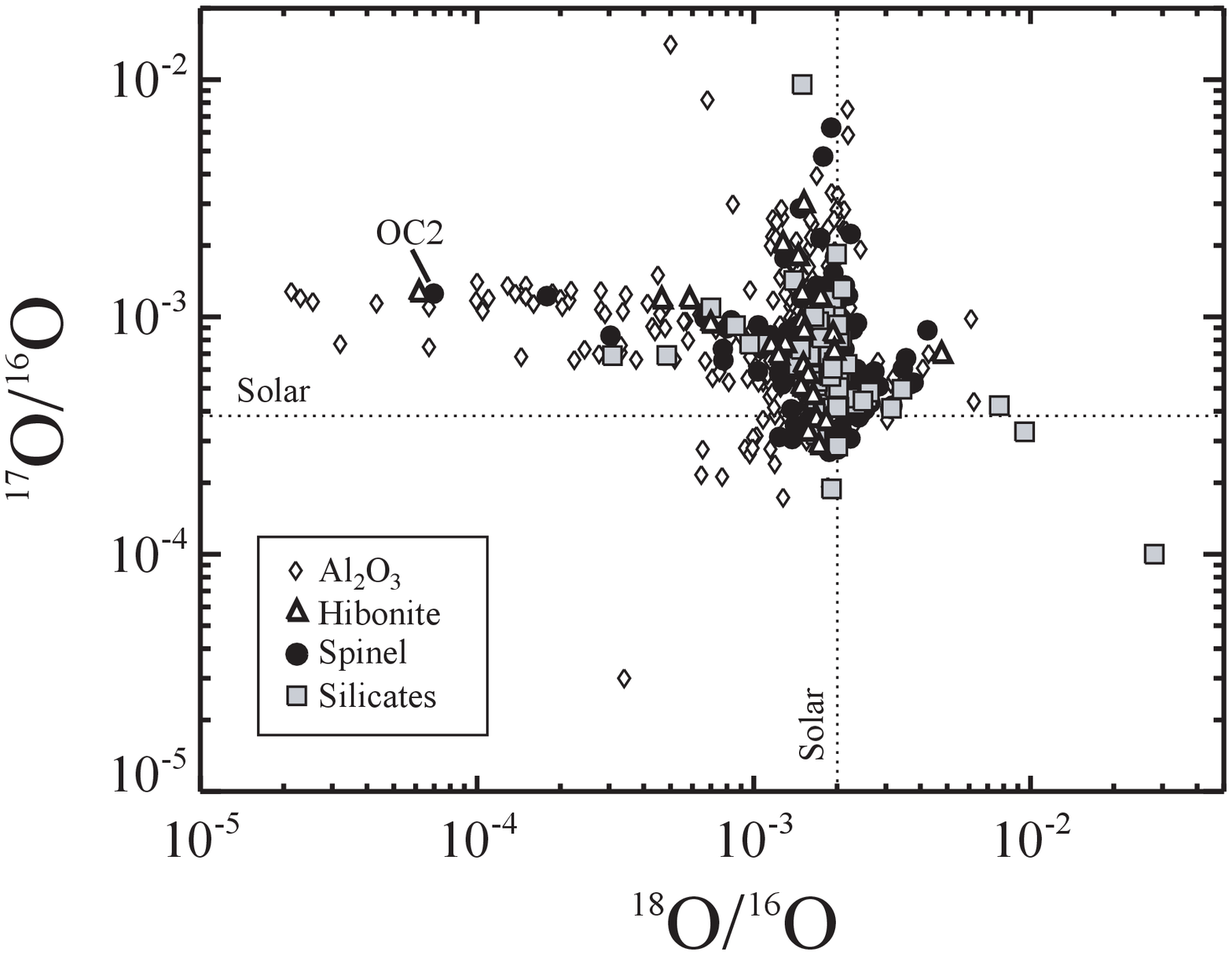}}
\caption{O isotopic ratios measured in presolar corundum, hibonite, spinel, and 
silicate grains 
\citep{nittlersolo:97,nittler:97,choi:98,messenger:03,zinner:03,mostefaoui:04,nguyen:04,nittler:05}.
Grain OC2 is a member of the rare Group~2 class of highly \ox{18}-depleted 
grains.} 
\label{fig:grains_o}
\end{figure}

\begin{table} 
\begin{center} 
\caption{The measured composition of presolar spinel grain 
OC2.\label{tab:OC2}} 
\begin{tabular}{crrr} 
\hline\hline 
 & OC2 & solar ratio \\
\hline
$\delta(^{25}$Mg/$^{24}$Mg) & 433.0 $\pm$ 10.0 & 0.1266 \\ 
$\delta(^{26}$Mg/$^{24}$Mg) & 1170.0 $\pm$ 15.0 & 0.1393 \\ 
Al/Mg & 2.18 $\pm$ 0.39 & 0.079 \\
$^{17}$O/$^{16}$O & 1.25 $\pm$ 0.07 $\times 10^{-3}$ & 3.83 $\times 10^{-4}$ \\ 
$^{18}$O/$^{16}$O & 6.94 $\pm$ 1.34 $\times 10^{-5}$$^a$ & 2.00 $\times 10^{-3}$ \\ 
$\delta(^{57}$Fe/$^{56}$Fe) & 170 $\pm$ 191 & 0.0231 \\ 
$\delta(^{50}$Cr/$^{52}$Cr) & 26 $\pm$ 71 & 0.0519 \\ 
$\delta(^{53}$Cr/$^{52}$Cr) & -56 $\pm$ 45 & 0.1135 \\ 
$\delta(^{54}$Cr/$^{52}$Cr) & 102 $\pm$ 117 & 0.0282 \\ 
\hline 
\end{tabular}
\end{center} 
{\footnotesize  All errors are given at 1$\sigma$.\\ $^a$Quoted here is the experimental 
statistical uncertainty, which does not take into account the possible effect of pollution of
terrestrial material discussed in Sec.~\ref{sec:comparison}.}
\end{table}

The O isotopic composition of grain OC2 (Fig.~\ref{fig:grains_o}) is also quite extreme with respect to 
typical compositions of presolar oxide grains. While its \ox{17}/\ox{16} ratio is enriched by a 
factor of 3.3 relative to solar, its $^{18}$O/$^{16}$O ratio is at least 26 times smaller than that of 
the Solar System. Presolar oxides with large \ox{18} depletions have been classified as ``Group~2'' 
grains by \citet{nittler:94,nittler:97}, but these are in fact quite rare. Of 
some 600 known presolar corundum, hibonite (CaAl$_{12}$O$_{19}$) and spinel grains, only 10 have 
\ox{18}/\ox{16} ratios lower than 10$^{-4}$, similar to that of OC2 (Fig.~\ref{fig:grains_o}). 

The Cr isotopic ratios were also measured in grain OC2, as well as the $^{57}$Fe/$^{56}$Fe ratio. 
From Table ~\ref{tab:OC2} we see that these values are typically
solar within the error bars.

The large \mg{25} and \mg{26} excesses observed in OC2 led \citet{zinner:05} to propose for this 
grain an origin in an intermediate-mass (IM)-AGB star, with mass between 4 -- 7 \msun\ and underging hot 
bottom burning \citep{herwig:05}, instead of the usual low-mass origin assigned to the majority of the 
other presolar oxide grains.
As for the late evolutionary phases of most stellar types, strong mass-loss and dust formation are 
observed around massive AGB stars. In fact, IM-AGB stars are part of the family of OH/IR stars, i.e. O-rich 
AGB stars extremely bright in the infrared, and astronomical observations indicates that OH/IR stars are 
the second most important source of dust in the Galaxy \citep{alexander:97,whittet:92}.

During the AGB phase, the H and He-burning shells are activated alternately in the deep layers of the 
star while the extended H-rich envelope loses mass through strong stellar winds.  During each 
He-burning episode, the sudden thermal runaway triggers the whole region between the H- and the 
He-shell (He intershell) to become convective. After a thermal pulse (TP) quenches, the 
base of the convective envelope can sink into the He intershell (third dredge-up, TDU), thus carrying 
to the surface nuclear processed material \citep[see][for a review on AGB stars]{herwig:05}.

The nucleosynthesis occurring in IM-AGB stars is different from that occurring in LM-AGB
stars because the temperature during thermal pulses can exceed $\simeq$ 350 
million degrees and thus the heavy Mg isotopes can be produced by $\alpha$-capture reactions on 
$^{22}$Ne, which is abundantly present in the He intershell due to the conversion of $^{14}$N from 
H-burning ashes into $^{22}$Ne by double $\alpha$ captures during the early phases of a thermal pulse 
\citep{karakas:06b}. The heavy Mg isotopes are then mixed to the envelope by the following TDU 
episode. In IM-AGB stars proton captures also occur at the base of the convective envelope (hot 
bottom burning, HBB). The MgAl chain is activated resulting typically in the destruction of 
$^{25}$Mg and the production of $^{26}$Al and $^{26}$Mg.  If the temperature exceeds $\simeq 80$ 
million degrees $^{24}$Mg also suffers proton captures \citep[see detailed discussion and models 
by][]{karakas:03,karakas:06b}. HBB also greatly affects the O isotopic composition, resulting in the 
production of $^{17}$O and the destruction of $^{18}$O, as observed in OC2. 
HBB can also prevent the surface of the star from becoming carbon-rich \citep{boothroyd:93}, 
by converting the dredged-up \car{12} to \nit{14}. 
Thus, IM-AGB stars might be expected to preferentially form 
oxide rather than carbonaceous phases like SiC and graphite, which require C$>$O. However, the 
formation of spinel cannot be completely ruled out for C/O ratios slightly above unity 
\citep[see discussion in][]{zinner:05}.

Stellar sources of dust in the Galaxy other than AGB stars \citep[see, e.g., Table 1 
of][]{alexander:97} seem unlikely to be the site of origin of grain OC2. The ejecta 
of supernova explosions are enriched in $^{25}$Mg, $^{26}$Mg and $^{26}$Al when considering solar 
metallicity models. However, contrarily to the composition of OC2, $^{16}$O and $^{18}$O are 
produced, while $^{17}$O is typically much depleted \citep{rauscher:02,limongi:03}. Wolf-Rayet stars, 
of which the WC type are also observed to generate dust \citep[see e.g][]{williams:87},
could also produce the O isotopic composition of grain OC2 together with excesses in $^{26}$Al at a time 
just 
before the transition from WN to WC occurs \citep{arnould:97}. However, no production of $^{25}$Mg is 
expected, at least until core He burning starts and the star moves into the WC/WO phases, at which 
point also large depletions of $^{17}$O and possible enhancements in $^{18}$O are predicted. Nova 
nucleosynthesis is predicted to produce high enhancements in $^{25}$Mg, $^{26}$Mg and $^{26}$Al, however, 
these are typically accompanied by high enhancements also in both $^{17}$O and $^{18}$O \citep{jose:04}. 
Only model CO1 of \citet{jose:04}, a CO nova of low mass 0.6 \msun, produces a deficit in $^{18}$O 
comparable to that observed in grain OC2. However, the $^{17}$O/$^{16}$O in this model is three 
times higher than in OC2, and the excesses in the heavy Mg isotopes are: 
$\delta(^{25}$Mg/$^{24}$Mg)=836 and $\delta(^{26}$Mg/$^{24}$Mg)=19062 
(without adding up the contribution of \al{26}, with \al{26}/\al{27}=0.006)! Even 
if mixing between this type of nova material and material of solar-like composition were invoked, it 
would be impossible to avoid producing $^{26}$Mg excesses much higher than those observed in 
grain OC2 at the time when the $^{25}$Mg/$^{24}$Mg ratio is matched.

The aim of this paper is to analyse the O, Mg, Al, Cr and Fe isotopic compositions predicted by detailed 
models of AGB stars of different masses and metallicities and discuss them in the light of the 
precise measurements of the composition of grain OC2. In this way, we can test the idea that grain 
OC2 originated in an IM-AGB star. There are few direct constraints available to test
theoretical models of these stars mostly owing to the fact that they are much rarer than
their lower-mass counterparts. There are observations available for Li, C, O, $^{12}$C/$^{13}$C and heavy 
elements abundances, in particular for AGB stars in the Magellanic Clouds \citep[see 
e.g.][]{wood:83,plez:93,smith:95,vanloon:99}.
The Li, C and O abundances and the $^{12}$C/$^{13}$C ratios have been used as tests for the occurrence 
of HBB. Stellar models are able to explain the fact that the majority of AGB stars of high luminosity are 
O rich, as well as to reproduce the observed low $^{12}$C/$^{13}$C 
ratios and high Li abundances \citep[e.g.][]{boothroyd:93,mazzitelli:99}. 

The paper is structured as follows: in Sect.~\ref{sec:models} we describe the numerical method and the 
structure features of the stellar models presented in this paper. In Sect.~\ref{sec:comparison} we compare  
the composition of grain OC2 to the compositions derived from our stellar models and discuss the model 
uncertainties. In Sect.~\ref{sec:concl} we outline our conclusions.

\section{Methods and models} \label{sec:models}

The evolution of stars of different masses and metallicities was computed
from the zero-age main sequence to near the tip of the AGB using the Monash
version of the Mt. Stromlo Stellar Structure code \citep[see][and references therein for 
details]{frost:96}. The models of IM-AGB stars that are used throughout this paper have 
masses of 5 \msun\ with $Z=0.02$ and 0.008 and 6.5 \msun\ with $Z=0.02$
and 0.012. The $Z = 0.02$ computations were assumed to have initial abundances taken
from \citet{anders:89}, whereas the $Z = 0.012$ models have initial abundances taken from 
\citet{asplund:05} for elemental abundances and \citet{lodders:03} for initial isotopic 
ratios. 
Most presolar grains recovered from meteorites have come from LM-AGB 
stars so we also present results from one low-mass low-$Z$ AGB model with 2.5 Msun
and $Z = 0.004$.  Models of AGB stars with $Z < 0.004$ (or $Z=0.004$ in the
case of IM-AGB stars) are not presented because presolar grains most likely
originated in field stars from the solar neighborhood. While we do not have any 
reliable observational information on interstellar dust lifetimes, theoretical estimates give 
values $<$ 1 Gyr \citep{jones:97}.

All models include mass loss on the AGB according to the prescription
of \citet{vassiliadis:93}, show deep TDU and with
the exception of the 2.5 \msun\ case, HBB, which prevents them from becoming C rich. For the 5 
\msun\, $Z=0.02$ model
we tested the effect of using the mass-loss rate from \citet{vanloon:05}. We find that the 
two formulas produce the same mass-loss rate in the final phases of the evolution 
(the last 8 out of 24 pulses), which is consistent given that the formula of \citet{vanloon:05} 
is valid for the superwind phase of extreme mass loss in O-rich AGB stars. 
In Table~\ref{tab:models} we present 
some features of the stellar models, where M5Z02 denotes the 5 \msun\ model
with $Z=0.02$. We include the number of TPs, the core mass at 
the first TP, the total amount of matter dredged-up into the envelope, 
$M_{\rm dred}^{\rm tot}$, and the maximum dredge-up efficiency, $\lambda$
\citep[for the definition, see][]{karakas:02}, where all masses are given in
solar units. 
We also present the maximum temperature at the base of the convective envelope, 
$T_{\rm BCE}^{\rm max}$, and in the He-intershell, $T_{\rm He}^{\rm max}$,
both in units of $10^{6}$ K; and the core and envelope mass at the final
time-step. In all IM-AGB models, HBB was shut off as a consequence
of mass loss. Once the envelope mass drops below about 1.5 \msun\
the temperature rapidly drops below that required to sustain proton
capture nucleosynthesis \citep{karakas:06b}. Dredge-up may continue, adding further 
\car{12} to the envelope and, depending on the initial metallicity, increase the C/O to $\ge$ 1 
\citep{frost:98,karakas:06b}. From \citet{karakas:06b}, however, this only occurs in IM-AGB stars 
with $Z \le 0.004$. 


\begin{table*}
\caption{Structural properties of the AGB models, see the text for details.\label{tab:models}} 
\begin{tabular}{rrrrrr}
\hline\hline
Models: & M5Z02 & M6.5Z02 & M6.5Z012 & M5Z008 & M2.5Z004 \\ 
\hline 
No. TP  &  24 & 40 & 51 &  59 & 34 \\
$M_{\rm c}(1)$ & 0.861 & 0.951 & 0.956 & 0.870 & 0.602 \\
$M_{\rm dred}^{\rm tot}$ & 5.027($-2$) & 4.696($-2$) & 6.483($-2$) & 1.745($-1$) & 1.869($-1$) \\
$\lambda_{\rm max}$ & 0.961 & 0.910 & 0.940 & 0.952 &  0.820 \\
$T_{\rm BCE}^{\rm max}$ & 64 & 87 & 90 & 81 & --  \\ 
$T_{\rm He}^{\rm max}$ & 352 & 372 & 370 & 366 & 308 \\
$M_{\rm env}^{\rm f}$ & 1.499 & 1.507 & 1.389 & 1.387 & 0.685 \\
$M_{\rm core}^{\rm f}$ & 0.874 & 0.963 & 0.967 & 0.886 & 0.673 \\
\hline
\end{tabular}
\end{table*}

We performed detailed nucleosynthesis calculations using a post-processing
code which includes 74 species, 506 reactions and time-dependent diffusive mixing 
in all convective zones \citep{cannon:93}. Details on the post-processing code and 
network are outlined in \citet{lugaro:04b} and \citet{karakas:06b} and will not be 
repeated here. Details on the reaction rates can be found in \citet{lugaro:04b}, 
where it is outlined which 
of the proton, $\alpha$ and neutron capture reaction rates we have updated 
\citep[from the REACLIB data tables][1991 version]{thielemann:86}, 
according 
to the latest experimental results. More recent updates include the $^{22}$Ne $+ \alpha$ 
reaction rates \citep{karakas:06b} and the NACRE rates \citep[][NACRE]{angulo:99} for the proton 
captures rates of the NeNa and MgAl chains. The case of the $^{17}$O($p,\alpha$)$^{14}$N reaction, 
of particular importance for the present work, is discussed in Sect.~\ref{sec:rates}.

For the 5 \msun\ $Z=0.008$ and 2.5 \msun\ $Z=0.004$ models, we assume an initial 
$\alpha$-enhanced mixture for the elements O, Ne, Mg, Si and S typical of thin-disk stars 
\citep{reddy:03}. The initial composition of the O and Mg isotopes are determined by the 
$\alpha$-enhancement of $^{16}$O and $^{24}$Mg, so that [O/Fe] $= +0.4$ and [Mg/Fe] $= +0.27$ at [Fe/H] $= 
-$1, while the abundances of the neutron-rich isotopes are scaled with the initial metallicity. 
Thus, the $^{17}$O/$^{16}$O and $^{18}$O/$^{16}$O ratios are lower than their solar values 
(see Table~\ref{tab:OC2}), by factors 1.44 and 1.91 
at $Z=0.008$ and 0.004, respectively. 
With this choice we obtain, at any given metallicity, 
higher O ratios than suggested by \citet{timmes:95} by following the Galactic chemical evolution,
and where the O ratios are scaled with the metallicity. However, there are still large 
uncertainties in the evolution of the O isotopic ratios with time \citep{prantzos:96,romano:03}.
Most importantly for the discussion here, the effect of the first and second dredge-up mixing
events and HBB almost completely erases any record of the initial O ratios at the surface
during the TP-AGB phase. 

\section{Comparison of model predictions and the composition of OC2} \label{sec:comparison}

\begin{figure}
\resizebox*{\hsize}{!}{\includegraphics[angle=270]{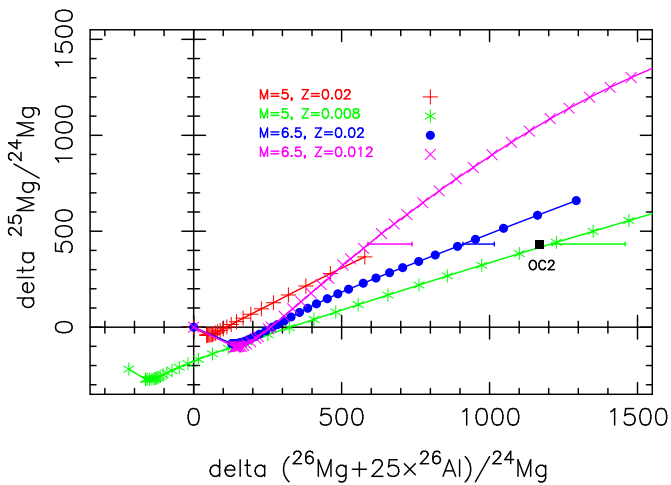}}
\caption{The Mg isotopic compositions of grain OC2 is compared to our models 
of IM-AGB stars. The 2$\sigma$ uncertainties for OC2 are roughly within the symbol. Each 
symbol for model predictions represents the composition after a TDU episode. 
As indicated in the x-label the $\delta(^{26}$Mg/$^{24}$Mg) measured in OC2 is compared 
to prediction lines calculated by including the abundance of $^{26}$Al multiplied by a factor of 
25 to take into account the fact that Al is preferentially included in spinel by such factor (see 
text). The uncertainty range in the predictions derived from the calculation of this factor are 
represented by the error bars connected to each prediction line at $\delta(^{25}$Mg/$^{24}$Mg)=433.
Solar composition is represented by ticked axis at $\delta=0$.}
\label{fig:selectMg}
\end{figure}

In Fig.~\ref{fig:selectMg} the Mg isotopic composition of grain OC2 is compared to that predicted by our 
calculations of IM-AGB models. The predictions shown in this plot are calculated by 
including the abundance of $^{26}$Al multiplied by 25, taking into account the 
fact that Al was incorporated in spinel grains during their formation approximately 25 times more 
preferentially than Mg, given that stoichiometric spinel by definition is MgAl$_{2}$O$_{4}$, i.e. it has
Al/Mg=2, while this ratio is 0.079 in the Solar System \citep{zinner:05}. We have explored the 
uncertainty related to this $^{26}$Al multiplication factor by considering the Al/Mg ratios predicted at 
the stellar surface in our models at the time when $\delta(^{25}$Mg/$^{24}$Mg)$_{\rm OC2}$ is matched, 
which are 5\% to 20\% lower than in the Solar System, and the error bars of the Al/Mg ratio measured in 
OC2 (see 
Table~\ref{tab:OC2}). The resulting uncertainties are represented by the error bars connected to the 
prediction lines in Fig.~\ref{fig:selectMg}.

First, let us consider the 5 \msun\ $Z=0.02$ model, which does not reach the excess in $^{25}$Mg shown 
by OC2. The $^{25}$Mg/$^{24}$Mg ratio is determined by the combined effect of the He 
intershell temperatures and the total amount of material mixed to the surface by the TDU.
An increase in either of those two quantities would lead to 
higher $^{25}$Mg/$^{24}$Mg ratios, so the OC2 data indicate that an IM-AGB parent star of this grain must 
have 
experienced either temperatures higher than 352 million degrees, or a TDU mass higher than 0.05 \msun, or 
both. Second, it appears that a narrow range of HBB 
temperatures is required in order to produce enough $^{26}$Al to match the excess at mass 26 shown 
by OC2 at the given $\delta(^{25}$Mg/$^{24}$Mg) value, avoiding an increase of the 
$\delta(^{25}$Mg/$^{24}$Mg) value itself. This occurs in the models of 5 \msun\ and $Z=0.008$ 
metallicity and 6.5 \msun\ and $Z=0.02$ metallicity (in this latter case within reaction rate 
uncertainties, see Sect.~\ref{sec:rates}), where the maximum temperature at the base of the 
convective envelope reaches 81 and 87 million degrees, respectively. For the 5 \msun\ $Z=0.02$ case 
the temperature is too low, $\leq$ 64 million degrees, to produce $^{26}$Al, while in the 6.5 \msun\ 
$Z=0.012$ case the temperature is too high, $\leq$ 90 million degrees, so that also $^{25}$Mg is 
produced by HBB. In principle, it is possible that for temperatures somewhat higher than 90 million 
degrees one could achieve again the 
composition of OC2 via HBB, as more $^{25}$Mg is converted into $^{26}$Al. However, these 
temperatures are not achieved in our models, and, moreover, they would pose a  
problem to match the $^{17}$O/$^{16}$O ratio of the grain (see discussion below and in 
Sect.~\ref{sec:rates}).

Note that the $\delta(^{25}$Mg/$^{24}$Mg) and $\delta(^{26}$Mg/$^{24}$Mg) values for the 5 \msun\ 
$Z=0.008$ model reach values higher than 2000 and 5000, respectively, outside the 
range shown in Fig.~\ref{fig:selectMg}. (For the 6.5 \msun\ $Z=0.012$ model, also outside the 
range of the figure, $\delta$-values reach $\simeq$2000 and $\simeq$2500, respectively). This 
could represent a difficulty for matching the composition of grain OC2 using the 5 \msun\ $Z=0.008$ 
model. It is much more likely for a grain to be produced with a composition that corresponds to 
abundances in the envelope at the end of the AGB. This is because radio and infrared observations 
confirm that mass-loss rates of AGB stars increase with time and that these stars expel a large part of 
their envelope towards the end of their evolution via a strong {\it superwind}, with $\dot{M} \sim$ few 
$10^{-4}$ \msun/yr \citep[e.g.][]{vassiliadis:93,vanloon:05}, leading to the formation 
of a planetary nebula. In the case of our 5 \msun\ $Z = 0.008$ model, 
about 40\% of the mass is lost in the last few TPs when the Mg isotopic ratios are well above that 
observed in OC2. Instead, the mass lost in between the TDU episodes that cover the composition of OC2 
is only $\simeq$ 0.005 \msun, hence, the probability of a grain forming with such composition is 
much smaller, $\simeq$ 0.1\%, even if this occurrence cannot be ruled out. From this point of view, 
the 6.5 \msun\ $Z=0.02$ model could be favored for the parent star of OC2, since it reaches the 
observed $\delta(^{25}$Mg/$^{24}$Mg) towards the end of its evolution and the $\delta(^{26}$Mg/$^{24}$Mg) 
value is within reaction rate uncertainties (see Sect.~\ref{sec:rates}).
However, it should be also kept in mind that the mass-loss prescription is one of the largest 
uncertainties in AGB models, as discussed in Sect.~\ref{sec:uncert}. 

\begin{figure}
\resizebox*{\hsize}{!}{\includegraphics[angle=270]{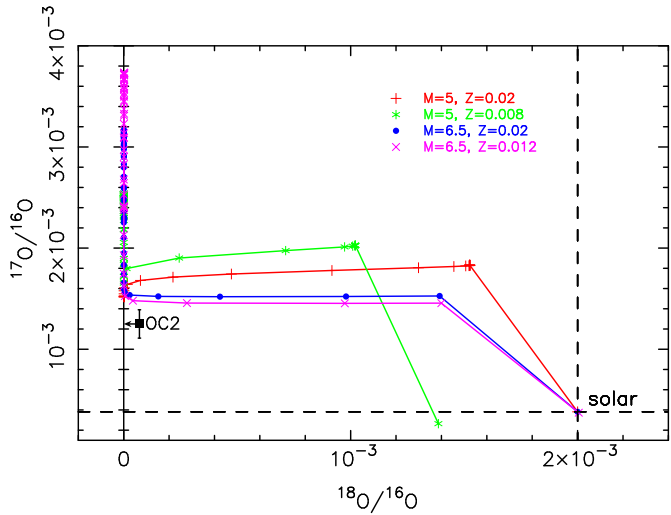}}
\caption{The O isotopic compositions of grain OC2 is compared to predictions from our IM-AGB stars. 
The 2$\sigma$ uncertainties 
for the $^{17}$O/$^{16}$O ratio of OC2 are indicated by the error bar, while for the 
$^{18}$O/$^{16}$O ratio they are roughly within the symbol. The arrow indicates the effect of possible 
contamination of terrestrial material, as discussed in the text. Solar ratios are indicated by 
long-dashed lines.}\label{fig:selectO}
\end{figure}

\begin{figure}
\resizebox*{\hsize}{!}{\includegraphics[angle=270]{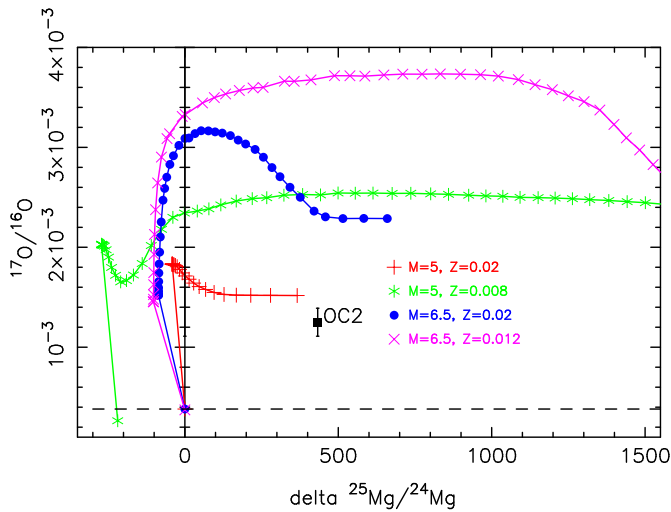}}
\caption{The $^{17}$O/$^{16}$O ratio is plotted as function of the 
$\delta(^{25}$Mg/$^{24}$Mg) for our IM-AGB models.}
\label{fig:Mg25O17}
\end{figure}

In Fig.~\ref{fig:selectO}, the O isotopic composition of grain OC2 is compared to that predicted by 
our IM-AGB models, and in Fig.~\ref{fig:Mg25O17} the $^{17}$O/$^{16}$O ratio 
is plotted as function of the $\delta(^{25}$Mg/$^{24}$Mg) value. These figures show that the 
$^{17}$O/$^{16}$O ratio of grain OC2 is not matched by any of the models, as they always produce too 
high a ratio. As discussed by \citet{nollett:03}, the $^{17}$O/$^{16}$O equilibrium ratio 
during proton captures is mainly determined by the ratio of the rates of the nuclear reactions that produce
and destroy $^{17}$O, i.e. the $^{16}$O($p,\gamma)^{17}$F and the $^{17}$O($p,\alpha)^{14}$N
reactions, respectively. This ratio reaches a minimum of $\simeq$ 0.0011 around 50 million degrees and then 
increases again for higher temperatures reaching 0.008 at 100 million degrees \citep[see Fig. 8 
of][]{nollett:03}. During HBB, the 
$^{17}$O/$^{16}$O equilibrium ratio is reached at the base of the convective envelope and the envelope 
material is efficiently 
replaced by material with the $^{17}$O/$^{16}$O equilibrium ratio after $\sim$ 5 TPs. 
The 5 \msun\ $Z=0.02$ model comes closest to producing the 
needed ratio because of its lower temperature, however, this model is the furthest from matching 
the measured $\delta(^{26}$Mg/$^{24}$Mg) value. The high temperatures at which HBB operates in the 5 
\msun\ $Z=0.008$ and in the 6.5 \msun\ $Z=0.02$ models lead to an even worse match with the O 
isotopic composition of grain OC2. The 6.5 \msun\ $Z=0.012$ model follows the expected trend as the 
$^{17}$O/$^{16}$O is the highest in this case. 

As for the $^{18}$O/$^{16}$O ratio, because of HBB, all the models reach ratios of the order of 
10$^{-6}$ - 10$^{-7}$, much lower than that shown by grain OC2, after only about 5 to 15 thermal 
pulses, much earlier than when the required Mg isotopic composition is reached. So, it is not possible to 
match the observed value by any of our models, 
but we note that there is always surface contamination on sample mounts and residual oxygen 
in the ion microprobe vacuum system. The very low measured \ox{18}/\ox{16} ratio for grain OC2 was 
based on 35 actual counted \ox{18} atoms. If the grain actually had an \ox{18}/\ox{16} ratio of zero, 
this low measured \ox{18} signal would correspond to a 2\% level of terrestrial contamination, 
which is perfectly reasonable. Thus, we consider it likely that the true \ox{18}/\ox{16} ratio of 
OC2 was indeed lower, and do not consider the mis-match with HBB models to be a major problem.
 
 
\subsection{The effect of varying the reaction rates} \label{sec:rates}

\begin{figure}
\resizebox*{\hsize}{!}{\includegraphics[angle=270]{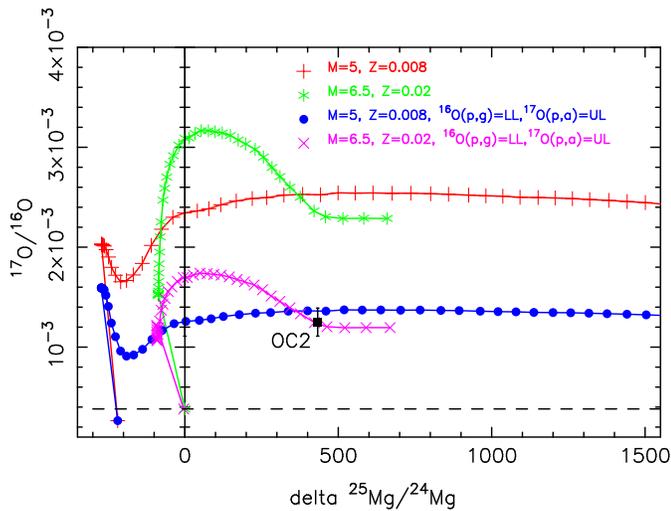}}
\caption{Selected IM-AGB model predictions from Fig.~\ref{fig:Mg25O17} are 
shown, with extra model predictions calculated using the lower limit (LL) and the upper limit (UL) 
for the $^{16}$O($p,\gamma)^{17}$F and the $^{17}$O($p,\alpha)^{14}$N reaction rates, respectively.} 
\label{fig:o16pgo17pa}
\end{figure}

As discussed above, the $^{17}$O/$^{16}$O ratio only depends on the ratio of the 
$^{16}$O($p,\gamma)^{17}$F and the $^{17}$O($p,\alpha)^{14}$N reaction rates, so it is important to 
carefully evaluate the uncertainties related to these reaction rates in the temperature range of 
HBB, $\simeq$ 60 to 100 million degrees. The $^{17}$O($p,\alpha)^{14}$N reaction rate in this 
temperature range is almost completely determined by a resonance at 70 keV. In the models presented 
above we have used the rate from \citet{blackmon:95} and \citet{landre:90}, which is the same as in 
the NACRE compilation. We have also calculated a new rate for this reaction on the basis of the latest 
available experimental information \citep{chafa:05,fox:05}. We have found a rate close to  
NACRE (within a few percent, for the temperature of interest in this study) with an uncertainty 
range of $\simeq +$ 25\% and $-$30\%, compared to the NACRE ranges of $\simeq +$ 33\% and $-$22\%. 
The $^{16}$O($p,\gamma)^{17}$F rate is mostly determined by direct capture and 
the astrophysical {\it S}-factor at zero energy is extrapolated from the available data using a given 
potential model \citep{angulo:99}. At low energy the experimental data show large error bars and the 
NACRE compilation adopts a 30\% uncertainty for the derived $S$-factor. This results in lower and 
upper limits for the rate a factor 1.43 lower and 1.30 higher, respectively, than the recommended 
value at the temperature of interest. As shown in Fig.~\ref{fig:o16pgo17pa}, the use of the upper limit 
for the $^{16}$O($p,\gamma)^{17}$F rate together with the lower limit for the
$^{17}$O($p,\alpha)^{14}$O rate is predicted to provide a good match to OC2, using 
the same models (5 \msun, $Z=0.008$ and 6.5 \msun, $Z=0.02$) that match its Mg isotopic composition.

The Mg isotopic composition is affected by the uncertainties in the $^{22}$Ne$+\alpha$ reaction 
rates as well as proton-capture reaction rates. The uncertainties of the $^{22}$Ne$+\alpha$ reaction rates 
have been 
recently re-evaluated by \citet{karakas:06b} and are relatively small (of the order
of $\simeq$60\%), while the 
uncertainties on the $^{25}$Mg($p,\gamma$)$^{26}$Al and $^{26}$Al($p,\gamma$)$^{27}$Si reaction rates are 
large (a factor of 3 and a factor of 1200, respectively) and affect, in 
particular, the production of $^{26}$Al. If we apply the uncertainties found by 
\citet{izzard:06} to the 
$^{26}$Al surface abundances of our models we find that the lower limit for the $^{26}$Al yield, 
is incompatible with the composition of OC2, while the
upper limit for the $^{26}$Al yield
would allow up to $\simeq$ 50\% more $^{26}$Al production. Combining this effect with the uncertainties of 
the 
$^{22}$Ne$+\alpha$ reaction rates (i.e. using the upper limit for the $^{22}$Ne($\alpha,\gamma$)$^{26}$Mg 
and the lower limit for the $^{22}$Ne($\alpha,n$)$^{25}$Mg reaction rates, respectively), we find that 
our 6.5 \msun, $Z=0.02$ model can also provides a match to the Mg isotopic composition of OC2.

As a final remark we note that our stellar structure calculations have been performed using a 
$^{14}$N($p,\gamma)^{15}$O reaction rate very similar to NACRE. However, direct experimental 
data are now available for this rate down to the typical H-burning temperatures of AGB stars 
\citep{runkle:05,luna:06}. They give a rate 40\% lower than NACRE. This has been shown to 
have an impact on the structure of AGB stars, in particular \citet{weiss:05} have shown that 
for a 5 \msun\ star of solar metallicity the peak luminosity in thermal pulses is higher, and the 
interpulse duration longer, while \citet{herwig:04c,herwig:06} have shown that for a 2 \msun\ star of 
$Z=$0.01 the choice of the new $^{14}$N($p,\gamma)^{15}$O reaction rate results in a more 
efficient third dredge-up. Further analysis is needed to test the effect of this updated reaction rate on 
nucleosynthesis in IM-AGB stars.

\subsection{The effect of model uncertainties} \label{sec:uncert}


One of the main uncertainties in AGB models is the choice of mass loss: a definitive agreement on 
the description of this phenomena is still missing. \citet{ventura:05a} calculated models with different 
choices of mass loss and showed that a stronger mass loss leads to less efficient HBB nucleosynthesis. In 
this case it may be more difficult to produce enough $^{26}$Al to match the $\delta(^{26}$Mg/$^{24}$Mg) of 
OC2.

Another uncertainty in AGB star models relates to the possible mixing of protons from the H-rich 
envelope into a tiny region at the top of He 
intershell (partial mixing zone, PMZ) at the end of each TDU episode, leading to the formation of a 
$^{13}$C pocket. This mixing is needed in order to reproduce the enhancements observed in AGB stars 
of heavy elements produced by $slow$ neutron captures (the $s$ process), with neutrons released by 
the $^{13}$C($\alpha$,n)$^{16}$O reaction \citep{gallino:98}. 
In the $^{13}$C pocket, the Mg isotopic composition  
can be altered by neutron captures. However, for IM-AGB models, we previously demonstrated 
\citep{karakas:06b} that the presence of a $^{13}$C pocket does not change the Mg isotopic ratios at 
the stellar surface because a reasonable value for the mass of the pocket is too small, $\simeq 
10^{-4}$ \msun, to produce any effect on the overall Mg nucleosynthesis. 

Other important uncertainties in the 
stellar modeling regard how convection and convective borders are treated.
\citet{ventura:05b} use a ``Full Spectrum of 
Turbulence'' (FST) prescription for convective regions, while we use Mixing Length Theory (MLT) with 
$\alpha$=1.7. The FST approach leads to a smaller efficiency of the TDU and higher HBB temperatures. 
We cannot yet make a direct comparison 
with these models as detailed nucleosynthesis is only available at the moment for stars 
of $Z=0.001$ with FST. As discussed by \citet{herwig:05} 
gravity waves or convective extra mixing at the base of the envelope and at the boundaries of the thermal 
pulse can affect the amount of TDU mass and also the temperature in the intershell. 
\citet{herwig:04a,herwig:04b} employs a special scheme to allow for some time-dependent 
hydrodynamical overshoot at convective boundaries. He typically finds very deep TDU and also that the 
convective thermal pulse penetrates inside the C-O degenerate core. 
Models are available for 
$Z=0.0001$, well below the metallicities we are considering here. Also in this case it remains to be 
analysed in detail if these models could provide a match to the composition of grain OC2.

Finally, we conclude our discussion of model uncertainties by noting that rotation and magnetic 
fields are not included in our models. These typically affect the 
occurrence of extra-mixing phenomena and have been studied for stars of masses lower \citep[see 
e.g.][]{denissenkov:03,talon:05}, or higher \citep[see e.g.][]{heger:00,maeder:05,yoon:05} than the 
IM star we are considering here. To our knowledge, however, there are no published models of massive AGB 
stars with rotation. Rotation and magnetic fields are not 1D phenomena and 
simplifications and parameterizations are required to input these physics into 1D stellar structure codes. 
These simplifications will result in extra parameters and uncertainties above
those already included in the simulations. For these reasons it is also valuable
to compute models without rotation, even though the effects of these phenomena on the evolution of massive 
AGB stars should be studied in the future.

\subsection{The case for a low-mass low-metallicity AGB star} \label{sec:lowmass}

\begin{figure}
\resizebox*{\hsize}{!}{\includegraphics[angle=270]{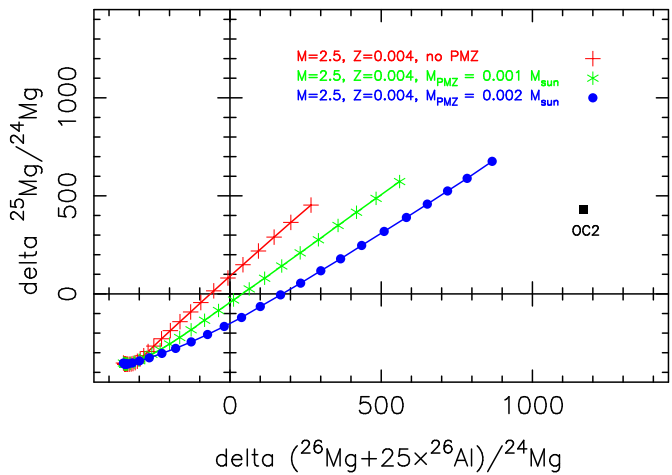}}
\caption{The Mg predictions are shown for a 2.5 \msun\ $Z=0.004$ model 
with different choices of the extension in mass of the partial mixing zone 
(PMZ).}
\label{fig:m2.5z004}
\end{figure}

A LM-AGB star of mass $\simeq$ 2.5 to 3.5 \msun\ and low 
metallicity, $Z \simeq 0.004$ to $\simeq$ 0.008 can also produce the $^{25}$Mg excess shown by OC2 
because the He intershell temperature and the TDU efficiency are high enough in these stars.  As 
an example, in Fig.~\ref{fig:m2.5z004} we present different models for a 2.5 \msun\ star of 
$Z$=0.004. As in the other figures, the predictions presented in this plot are calculated by 
including the abundance of $^{26}$Al multiplied by the chosen factor of 25, however 
in these models the contribution at mass 26 comes mostly from $^{26}$Mg.
In two runs of the post-processing of this model we artificially included the mixing of protons 
into the He intershell at the end of each TDU episode 
and we tested two different values for the extent in mass of the PMZ: 0.001 and 0.002 
\msun. More $^{26}$Mg is produced when the pocket is included because of neutron captures. 
Since this star is just not hot enough for HBB to occur we have to assume that some extra-mixing process 
at the base of the convective envelope, such as the cool bottom processing (CBP) studied by 
\citet{nollett:03}, is at work during the AGB phase, so that enough $^{26}$Al is produced and the 
observed $\delta(^{26}$Mg/$^{24}$Mg) is matched.
In order to reproduce the $\delta(^{26}$Mg/$^{24}$Mg) of OC2 a $^{26}$Al/$^{27}$Al ratio in the 
range 0.06 to 0.036 is required, which can be produced with CBP temperatures $\simeq$ 54 million K 
\citep{nollett:03} 
The low $^{18}$O/$^{16}$O ratio in OC2 is consistent with efficient CBP and the $^{17}$O/$^{16}$O ratio 
would be 0.00115 
at the CBP temperature derived from the $^{26}$Al/$^{27}$Al ratio.
Detailed calculations for CBP at low metallicities are required to 
test our present analysis before any strong conclusion can be drawn. We plan to include this 
process in our nucleosynthesis code in the future.



\subsection{The Fe and Cr isotopic composition} \label{sec:fecr}

In AGB stars the Fe isotopic composition is modified because the heavy Fe isotopes are produced by neutron 
captures occurring during the convective thermal pulses when the $^{22}$Ne($\alpha,n$)$^{25}$Mg 
neutron source is activated. This production is rather insensitive to the neutron 
captures occurring in the $^{13}$C pocket \citep{lugaro:04a}. The $\delta(^{57}$Fe/$^{56}$Fe) value,
at the time when $\delta(^{25}$Mg/$^{24}$Mg)$\simeq$433, is $\simeq$ 80 in our IM-AGB models and $\simeq$ 
370 in the 2.5 \msun\ $Z=0.004$ model. 
There are a couple of reasons for this difference: first, the integrated 
neutron flux increases with decreasing the metallicity as less material means that more free neutrons 
are available. Second, the dilution factor of the intershell material carried to the envelope by the 
TDU is about a factor of two higher in IM- than in LM-AGB stars. 
Unfortunately, the large uncertainty of the measured $\delta(^{57}$Fe/$^{56}$Fe) prevents us from 
determining which models represent the best match.

We cannot directly compare the Cr isotopic composition as we do not have the Cr isotopes in the network 
used to 
compute the present models. We can draw some general features on Cr isotopes in AGB stars considering 
Fig.~2 of \citet{lugaro:04a}, where results are presented for LM-AGB models of solar metallicity. 
The $\delta(^{50}$Cr/$^{52}$Cr) and $\delta(^{53}$Cr/$^{52}$Cr) are barely modified by neutron 
fluxes in AGB stars while the $\delta(^{54}$Cr/$^{52}$Cr) values are slightly higher than the 
$\delta(^{57}$Fe/$^{56}$Fe) values for a given stellar model. 
Since we are in the process of extending our network \citep{karakas:06a}, using preliminary 
models we have checked that these results also apply to our IM-AGB models. In fact, for the 5 \msun\ 
$Z=0.02$ model we obtain no changes in $\delta(^{50}$Cr/$^{52}$Cr) and $\delta(^{53}$Cr/$^{52}$Cr), while  
$\delta(^{54}$Cr/$^{52}$Cr)=170, compared to $\delta(^{57}$Fe/$^{56}$Fe)=71. For the 6.5 \msun\ $Z=0.012$ 
model we obtain $\delta(^{54}$Cr/$^{52}$Cr)=40, at pulse number 25. Our IM-AGB models would match the 
$\delta(^{54}$Cr/$^{52}$Cr) value measured in OC2, while the 2.5 \msun\ $Z=0.004$ model would produce 
$\delta(^{54}$Cr/$^{52}$Cr) values close to the 2$\sigma$ upper limit of OC2. 

Finally, we note that in low metallicity stars the 
initial $\delta(^{i}$Cr/$^{52}$Cr) could be largely negative because of the effect of the Galactic 
Chemical Evolution of the Cr isotopes \citep[see discussion in][]{zinner:05}.
This represents a hint against a LM-AGB star of low 
metallicity as the parent star for OC2, even though detailed models for the evolution of the Cr 
isotopes in the Galaxy should be performed to confirm this analysis. 

\section{Conclusions}\label{sec:concl}

We have shown that the peculiar isotopic composition of O, Mg and Al in presolar spinel grain OC2 could be 
the signature of an AGB star of intermediate mass and metallicity close to solar (roughly higher than 
0.008) suffering TDU and HBB, or of an AGB star of low mass and low metallicity (roughly lower than 
0.008) suffering TDU and very efficient CBP. While HBB occurs in our IM-AGB models, we do not 
simulate any extra-mixing in the LM-AGB model. Thus, the LM-AGB origin for grain OC2 has still to be 
tested against detailed models including CBP in low metallicity AGB stars. 
The large uncertainty in the Fe isotopic composition of OC2 does not allow us to determine
which model better represents the parent star of OC2, but the Cr isotopic
composition favors an origin in an IM-AGB star of metallicity close to solar.
In this case, the model conditions to reproduce the composition of OC2 are well defined:
a TDU mass $>$ 0.05 \msun\ and/or a maximum He-intershell temperature $>$ 360 million degrees, and a 
temperature at the base of the convective envelope in the range $\simeq$ 80 to 85 million degrees. These 
conditions are satisfied by our 5 \msun, $Z=0.008$ and 6.5 \msun, $Z=0.02$ models. Within this solution, we 
predict that the $^{16}$O($p,\gamma)^{17}$F and the $^{17}$O($p,\alpha)^{14}$N reaction rates should be 
close to their lower and upper limits, respectively. It remains to be seen if the proposed 
rate changes are consistent with other constraints, such as future nuclear expriments, new 
grains similar to OC2, or the production of $^{17}$O in other stellar types.

We note that using Salpeter's initial mass function, stars with mass from 5 to 7 \msun\ represent 
$\simeq$ 5\% of all stars. Thus, one might expect such a proportion of 
presolar grains from IM- 
relatively to LM-AGB stars. In principle, any grain with $^{18}$O/$^{16}$O $<$ 10$^{-4}$ and 
$^{26}$Mg excesses is a candidate for an IM-AGB origin. 
There are 69 oxide grains in our database
found by measurements of both $^{17}$O/$^{16}$O and $^{18}$O/$^{16}$O (thus
representing an unbiased sample) and showing $^{26}$Mg excesses. Of these, three have 
$^{18}$O/$^{16}$O lower than
10$^{-4}$. Thus, at most $\simeq$ 4\% of presolar oxides appear to have 
even the possibility to have 
formed in IM-AGB stars. \citet{bernatowicz:05} discuss detailed calculations of graphite and
TiC grain growth in carbon stars. Their Fig.~8 indicates that larger grains form in lower mass stars. 
\citet{nuth:06} propose a mechanism for the growth of crystals in AGB stars and red 
giants and show that only in low-mass ($< 3$ \msun) stars it is possible to form the large crystals found 
in primitive meteorites. Most of the oxide grains with both O and Mg measurements are larger than 1 
micron, thus it is quite possible that our dataset is biased towards grains from low mass stars.

Further analysis of presolar spinel grains may identify additional 
grains with isotopic compositions similar to OC2 and give more precise measurements of their Fe and Cr 
isotopic compositions. This will provide the opportunity to test 
the findings of the present work as well as the possibility to constrain massive AGB models using 
presolar grains.

\begin{acknowledgements}
ML gratefully acknowledges the support of NWO through the VENI grant and thanks the members of the 
stellar group of Onno Pols at Utrecht University for discussion. All computations were 
performed on Canadian Institute for Theoretical Astrophysics's Mckenzie cluster which was funded 
by the Canada Foundation for Innovation and the Ontario Innovation Trust.
The work of LRN and CA was supported by NASA. This work was partially supported by the Australian 
Research Council. We thank the anonymous referee for a thorough report, which helped to improve 
the paper. Finally, we thank Maurizio Busso for the original suggestion that OC2 shows the signature of 
an IM-AGB origin.
\end{acknowledgements}

%
%
%
%
%
%
%

\end{document}